\documentclass[letterpaper, 12pt]{article}[2000/05/19]
\usepackage[english]{babel}
\usepackage{amsfonts,amsmath,amssymb,amsthm,latexsym,amscd,mathrsfs}
\usepackage{ifthen,cite}
\usepackage[bookmarksnumbered=true]{hyperref}

\hypersetup{pdfpagetransition={Split}}

\newcommand{\evenhead}{Author \ name}
\newcommand{\oddhead}{Article \ name}
\newcommand{\theArticleName}{Article \ name}

\newcommand{\FirstPageHeading}[1]{\thispagestyle{empty}%
\noindent\raisebox{0pt}[0pt][0pt]{\makebox[\textwidth]{\protect\footnotesize \sf }}\par}

\newcommand{\ArticleName}[1]{\renewcommand{\theArticleName}{#1}\vspace{-2mm}\par\noindent {\LARGE\bf  #1\par}}
\newcommand{\Author}[1]{\vspace{5mm}\par\noindent {\Large  #1\par} \par\vspace{2mm}\par}
\newcommand{\Address}[1]{\vspace{2mm}\par\noindent {\it #1} \par}
\newcommand{\Email}[1]{\ifthenelse{\equal{#1}{}}{}{\par\noindent {\rm E-mail: }{\it  #1} \par}}
\newcommand{\URLaddress}[1]{\ifthenelse{\equal{#1}{}}{}{\par\noindent {\rm URL: }{\tt  #1} \par}}
\newcommand{\EmailD}[1]{\ifthenelse{\equal{#1}{}}{}{\par\noindent {$\phantom{\dag}$~\rm E-mail: }{\it  #1} \par}}
\newcommand{\URLaddressD}[1]{\ifthenelse{\equal{#1}{}}{}{\par\noindent {$\phantom{\dag}$~\rm URL: }{\tt  #1} \par}}

\newcommand{\Abstract}[1]{\vspace{6mm}\par\noindent\hspace*{10mm}
\parbox{140mm}{\small {\bf Abstract.} #1}\par}
\newcommand{\Keywords}[1]{\vspace{3mm}\par\noindent\hspace*{10mm}
\parbox{140mm}{\small {\bf Key words:} \rm #1}\par}
\newcommand{\Classification}[1]{\vspace{3mm}\par\noindent\hspace*{10mm}
\parbox{140mm}{\small {\it 2000 Mathematics Subject Classification:} \rm #1}\vspace{3mm}\par}
\newcommand{\ShortArticleName}[1]{\renewcommand{\oddhead}{#1}}
\newcommand{\AuthorNameForHeading}[1]{\renewcommand{\evenhead}{#1}}

\long\def\@makecaption#1#2{
  \sbox\@tempboxa{\small \textbf{#1.}\ \ #2}%
  \ifdim \wd\@tempboxa >\hsize
    {\small \textbf{#1.}\ \ #2}\par \else
    \global \@minipagefalse
    \hb@xt@\hsize{\hfil\box\@tempboxa\hfil}%
  \fi \vskip\belowcaptionskip}

\def\numberwithin#1#2{\@ifundefined{c@#1}{\@nocounterr{#1}}{%
  \@ifundefined{c@#2}{\@nocnterr{#2}}{%
  \@addtoreset{#1}{#2}%
  \toks@\@xp\@xp\@xp{\csname the#1\endcsname}%
  \@xp\xdef\csname the#1\endcsname
    {\@xp\@nx\csname the#2\endcsname.\the\toks@}}}}
\def\E^#1{{\buildrel #1 \over\vee}}
\newtheorem{theorem}{Theorem}

{\theoremstyle{definition}

}

\begin{document}

\FirstPageHeading{V.I. Gerasimenko}

\ShortArticleName{Derivation of the Boltzmann equation}

\AuthorNameForHeading{V.I. Gerasimenko}

\ArticleName{Approaches to Derivation of the Boltzmann \\Equation
with Hard Sphere Collisions}

\Author{V.I. Gerasimenko$^\ast$\footnote{E-mail: \emph{gerasym@imath.kiev.ua}}}

\Address{$^\ast$Institute of Mathematics of NAS of Ukraine,\\
        \hspace*{0.5mm} 3, Tereshchenkivs'ka Str.,\\
        \hspace*{0.5mm} 01601, Kyiv, Ukraine}

\bigskip
\Abstract{In the paper the possible approaches to the rigorous derivation
of the Boltzmann kinetic equation with hard sphere collisions from underlying
dynamics are considered. In particular, a formalism for the description of
the evolution of infinitely many hard spheres within the framework of marginal
observables in the Boltzmann--Grad scaling limit is developed. Also we give
consideration to one more approach of the description of the kinetic evolution
of hard spheres in terms of a one-particle distribution function governed by
the non-Markovian generalization of the Enskog equation and the Boltzmann--Grad
asymptotic behavior of its non-perturbative solution is established.}

\bigskip

\vspace{2pc}
\Classification{35Q20; 35Q82; 47J35; 82C05; 82C40.}

\makeatletter
\renewcommand{\@evenhead}{
\hspace*{-3pt}\raisebox{-7pt}[\headheight][0pt]{\vbox{\hbox to \textwidth {\thepage \hfil \evenhead}\vskip4pt \hrule}}}
\renewcommand{\@oddhead}{
\hspace*{-3pt}\raisebox{-7pt}[\headheight][0pt]{\vbox{\hbox to \textwidth {\oddhead \hfil \thepage}\vskip4pt\hrule}}}
\renewcommand{\@evenfoot}{}
\renewcommand{\@oddfoot}{}
\makeatother

\newpage
\vphantom{math}

\protect\tableofcontents
\vspace{1.5cm}



\section{Introduction}
As is known the collective behavior of many-particle systems can be effectively described
within the framework of a one-particle marginal distribution function governed by the
kinetic equation which is a scaling limit of underlying dynamics \cite{CGP97}-\cite{Sp91}.

The statement of a problem of the derivation of the Boltzmann kinetic equation from the
Hamiltonian dynamics goes back to D. Hilbert \cite{H}. The sequential analysis of this
problem within the framework of the perturbation theory was realized by N.N. Bogolyubov
\cite{B46}. The approach to the derivation of the Boltzmann equation as a result of the
scaling limit of the BBGKY hierarchy (Bogolyubov--Born--Green--Kirkwood--Yvon) was originated
in H. Grad's work \cite{GH}.

At present significant progress in the problem solving of the rigorous derivation of the
Boltzmann equation for a system of hard spheres in the Boltzmann--Grad scaling limit is
observed \cite{C72}-\cite{V}. Nowadays these results were extended on many-particle
systems interacting via a short-range potential \cite{SR12},\cite{PSS}.

The conventional approach to the derivation of the Boltzmann equation with hard sphere
collisions from underlying dynamics (see \cite{CGP97}-\cite{Sp91} and references cited
therein) is based on the construction of the Boltzmann--Grad asymptotic behavior of a
solution of the Cauchy problem of the BBGKY hierarchy for hard spheres represented in
the form of the perturbation (iteration) series \cite{U01}.

Another approach to the description of the many-particle evolution is given within the
framework of marginal observables governed by the dual BBGKY hierarchy \cite{BGer}. One
of the aims of the present paper consists in the description of the kinetic evolution
of hard spheres in terms of the evolution of observables (the Heisenberg picture of the
evolution). The Heisenberg picture representation of dynamics is in fact the best
mathematically fully consistent formulation, since the notion of state in more subtle,
in particular it depends on the reference frame. For example, for this reason it is useful
in relativistic quantum field theory.

One more approach to the description of the kinetic evolution of hard spheres is based
on the non-Markovian generalization of the Enskog equation. In paper \cite{GG} in case
of initial states specified in terms of a one-particle distribution function the equivalence
of the description of the evolution of hard sphere states by the Cauchy problem of the BBGKY
hierarchy and in terms of a one-particle distribution function governed by the generalized
Enskog kinetic equation was established. The Boltzmann--Grad asymptotic behavior of a
non-perturbative solution of the Cauchy problem of the generalized Enskog equation is
described by the Boltzmann equation with hard sphere collisions \cite{GG04}. We note that
within the framework of the perturbation theory a particular case of this approach reduces
to the Bogolyubov's method \cite{B46} of the derivation of the Boltzmann equation and its
generalizations.

We remark also that the rigorous method of the description of the kinetic evolution of a
tracer hard sphere and an environment of hard spheres was developed in paper \cite{GG12}
and a rigorous derivation of the Brownian motion in the hydrodynamic limit was realized
in \cite{BGS}. The approaches to the rigorous derivation in a mean field scaling limit of
quantum kinetic equations from underlying many-particle quantum dynamics were considered
in reviews \cite{G09},\cite{G12}.

The paper is organized as follows.

In section 2 we formulate necessary preliminary facts about evolution equations of a system
of hard spheres and review recent rigorous results on the derivation of the Boltzmann kinetic
equation from underlying dynamics. In section 3 we develop an approach to the description of
the kinetic evolution of infinitely many hard spheres within the framework of the evolution
of marginal observables. For this purpose we establish the Boltzmann--Grad asymptotic behavior
of a solution of the Cauchy problem of the dual BBGKY hierarchy for marginal observables of
hard spheres. The constructed scaling limit is governed by the set of recurrence evolution
equations, namely by the dual Boltzmann hierarchy. Furthermore, in this section the relationships
of the dual Boltzmann hierarchy for the limit marginal observables with the Boltzmann kinetic
equation are established. In section 4 for the description of the kinetic evolution of hard
spheres we develop one more approach based on the non-Markovian generalization of the Enskog
kinetic equation. We prove that the Boltzmann--Grad scaling limit of a non-perturbative solution
of the Cauchy problem of the generalized Enskog kinetic equation is governed by the Boltzmann
equation and the property on the propagation of initial chaos is established. Moreover, the
Boltzmann--Grad asymptotic behavior of a one-dimensional system of hard spheres is outlined.
Finally, in section 5 we conclude with some observations and perspectives for future research.


\section{Evolution equations of many hard spheres}
It is well known that a description of many-particle systems is formulated in terms
of two sets of objects: observables and states. The functional of the mean value of
observables defines a duality between observables and states and as a consequence
there exist two approaches to the description of the evolution. Usually the evolution
of many-particle systems is described within the framework of the evolution of states
by the BBGKY hierarchy for marginal distribution functions. An equivalent approach
to the description of the evolution of many-particle systems is given in terms of
marginal observables governed by the dual BBGKY hierarchy.

\subsection{The BBGKY hierarchy for hard spheres}
We consider a system of identical particles of a unit mass with the diameter $\sigma>0$,
interacting as hard spheres with elastic collisions. Every hard sphere is characterized
by the phase space coordinates:
$(q_{i},p_{i})\equiv x_{i}\in\mathbb{R}^{3}\times\mathbb{R}^{3},\,i\geq1.$

Let $L^{1}_{\alpha}=\oplus^{\infty}_{n=0}\alpha^n L^{1}_{n}$ be the space of
sequences $f=(f_0,f_1,\ldots,f_n,\ldots)$ of integrable functions $f_n(x_1,\ldots,x_n)$
defined on the phase space of $n$ hard spheres, that are symmetric with respect
to permutations of the arguments $x_1,\ldots,x_n$, equal to zero on the set of the
forbidden configurations: $\mathbb{W}_n\equiv\{(q_1,\ldots,q_n)\in\mathbb{R}^{3n}\big|
|q_i-q_j|<\sigma\,\,\,\text{for at least one pair}\,\, (i,j):i\neq j\in(1,\ldots,n)\}$ and
equipped with the norm: $\|f\|_{L^{1}_{\alpha}}=\sum_{n=0}^{\infty}\alpha^n \|f_n\|_{L^{1}_n}
=\sum_{n=0}^{\infty}\alpha^n \int dx_1\ldots dx_n|f_n(x_1,\ldots,x_n)|$,
where $\alpha>1$ is a real number. We denote by $L_{0}^1\subset L^{1}_{\alpha}$
the everywhere dense set in $L^{1}_{\alpha}$ of finite sequences of continuously
differentiable functions with compact supports.

The evolution of all possible states of a system of a non-fixed, i.e. arbitrary but
finite, number of hard spheres is described by the sequence
$F(t)=(1,F_{1}(t,x_1),\ldots,F_{s}(t,x_{1},\ldots,x_{s}),\ldots)\in L^{1}_{\alpha}$
of the marginal ($s$-particle) distribution functions $F_s(t,x_1,\ldots,x_s),\,s\geq1$,
governed by the Cauchy problem of the weak formulation of the BBGKY hierarchy \cite{CGP97}:
\begin{eqnarray}
  \label{NelBog1}
    &&\hskip-8mm\frac{\partial}{\partial t}F_s(t)=\big(\sum\limits_{j=1}^{s}\mathcal{L}^\ast(j)+
       \epsilon^{2}\sum\limits_{j_{1}<j_{2}=1}^{s}
       \mathcal{L}_{\mathrm{int}}^\ast(j_{1},j_{2})\big)F_{s}(t)+\\
    &&+\epsilon^{2}\sum_{i=1}^{s}\,\int_{\mathbb{R}^{3}\times\mathbb{R}^{3}}dx_{s+1}
       \mathcal{L}^\ast_{\mathrm{int}}(i,s+1)F_{s+1}(t),\nonumber\\ \nonumber\\
  \label{eq:NelBog2}
    &&\hskip-8mmF_s(t)_{\mid t=0}=F_s^{0,\epsilon}, \quad s\geq1.
\end{eqnarray}
In the hierarchy of evolution equations (\ref{NelBog1}) represented in a dimensionless
form the coefficient $\epsilon>0$ is a scaling parameter (the ratio of the diameter
$\sigma>0$ to the mean free path of hard spheres) and, if $t\geq0$, the operators
$\mathcal{L}^{\ast}(j)$ and $\mathcal{L}_{\mathrm{int}}^{\ast}(j_1,j_{2})$ are defined
on the subspace $L_{n,0}^1\subset L_{n}^1$ by the formulas:
\begin{eqnarray}\label{aL}
    &&\hskip-7mm\mathcal{L}^\ast(j)f_n\doteq -\langle p_{j},
        \frac{\partial}{\partial q_{j}}\rangle f_n,\\
  \label{aLint}
    &&\hskip-7mm\mathcal{L}_{\mathrm{int}}^\ast(j_{1},j_{2})f_{n}
        \doteq\int_{\mathbb{S}_{+}^2}d\eta\langle\eta,(p_{j_{1}}-p_{j_{2}})\rangle
        \big(f_n(x_1,\ldots,q_{j_{1}},p_{j_{1}}^\ast,\ldots,\\
    &&q_{j_{2}},p_{j_{2}}^\ast,\ldots,x_n)\delta(q_{j_{1}}-q_{j_{2}}+\epsilon\eta)-
        f_n(x_1,\ldots,x_n)\delta(q_{j_{1}}-q_{j_{2}}-\epsilon\eta)\big),\nonumber
\end{eqnarray}
where the following notations are used: the symbol $\langle \cdot,\cdot \rangle$
means a scalar product, $\delta$ is the Dirac measure,
$\mathbb{S}_{+}^{2}\doteq\{\eta\in\mathbb{R}^{3}\big|\left|\eta\right|=1,\,
\langle\eta,(p_{j_{1}}-p_{j_{2}})\rangle\geq0\}$ and the pre-collision momenta
$p_{j_{1}}^\ast,p_{j_{2}}^\ast$ are determined by the expressions:
\begin{eqnarray}\label{momenta}
   &&p_{j_{1}}^\ast=p_{j_{1}}-\eta\left\langle\eta,\left(p_{j_{1}}-p_{j}\right)\right\rangle,\\
   &&p_{j_{2}}^\ast=p_{j_{2}}+\eta\left\langle\eta,\left(p_{j_{1}}-p_{j_{2}}\right)\right\rangle.\nonumber
\end{eqnarray}

The adjoint Liouville operator $\mathcal{L}^\ast_s=\sum_{i=1}^{s}\mathcal{L}^\ast(i)+
\epsilon^{2}\sum_{i<j=1}^{s}\mathcal{L}_{\mathrm{int}}^\ast(i,j)$ is an infinitesimal generator
of the group of operators of $s$ hard spheres: $S_{s}^\ast(t)\equiv S_{s}^\ast(t,1,\ldots,s)$,
which is adjoint to the group of operators $S_{s}(t)$ defined almost everywhere on the phase
space $\mathbb{R}^{3s}\times(\mathbb{R}^{3s}\setminus \mathbb{W}_s)$ as the shift operator
of phase space coordinates along the phase space trajectories of $s$ hard spheres \cite{CGP97}.
The adjoint group of operators $S_{s}^\ast(t)$ coincides with the group of operators of hard
spheres $S_{s}(-t)$ \cite{CGP97}.

In case of $t\leq0$ the generator of the BBGKY hierarchy with hard sphere collisions is determined
by the corresponding operator \cite{CGP97}.

If initial data $F(0)=(1,F_1^{0,\epsilon},\ldots,F_n^{0,\epsilon},\ldots)\in L^{1}_{\alpha}$ and
$\alpha>e$, then for arbitrary $t\in\mathbb{R}$ a unique non-perturbative solution of the Cauchy
problem (\ref{NelBog1}),(\ref{eq:NelBog2}) of the BBGKY hierarchy with hard sphere collisions
exists and it is represented by a sequence of the functions \cite{GRS04}:
\begin{eqnarray}\label{F(t)}
    &&\hskip-8mm F_{s}(t,x_1,\ldots,x_{s})=\sum\limits_{n=0}^{\infty}\frac{1}{n!}
        \int_{(\mathbb{R}^{3}\times\mathbb{R}^{3})^{n}}dx_{s+1}\ldots dx_{s+n}
        \mathfrak{A}_{1+n}(-t,\{Y\},X\setminus Y)F_{s+n}^{0,\epsilon},\quad s\geq1,
\end{eqnarray}
where the generating operator of the $n$ term of series expansion (\ref{F(t)})
is the $(n+1)th$-order cumulant of adjoint groups of operators of hard spheres:
\begin{eqnarray}\label{nLkymyl}
   &&\hskip-8mm \mathfrak{A}_{1+n}(-t,\{Y\},X\setminus Y)=
      \sum\limits_{\texttt{P}:\,(\{Y\},X\setminus Y)={\bigcup\limits}_i X_i}
      (-1)^{|\texttt{P}|-1}(|\texttt{P}|-1)!
      \prod_{X_i\subset \texttt{P}}S_{|\theta(X_i)|}(-t,\theta(X_i)),
\end{eqnarray}
and the following notations are used: $\{Y\}$ is a set consisting of one element
$Y\equiv(1,\ldots,s)$, i.e. $|\{Y\}|=1$, $\sum_\texttt{P}$ is a sum over all possible
partitions $\texttt{P}$ of the set $(\{Y\},X\setminus Y)\equiv(\{Y\},s+1,\ldots,s+n)$
into $|\texttt{P}|$ nonempty mutually disjoint subsets $X_i\in(\{Y\},X\setminus Y)$,
the mapping $\theta$ is the declusterization mapping defined by the formula:
$\theta(\{Y\},X\setminus Y)=X$. The simplest examples of cumulants (\ref{nLkymyl}) are
given by the following expansions:
\begin{eqnarray*}
   &&\mathfrak{A}_{1}(-t,\{Y\})=S_{s}(-t,Y),\\
   &&\mathfrak{A}_{2}(-t,\{Y\},s+1)=S_{s+1}(-t,Y,s+1)-S_{s}(-t,Y)S_{1}(-t,s+1).
\end{eqnarray*}

For initial data $F(0)\in L^{1}_{\alpha,0}\subset L^{1}_{\alpha}$ sequence (\ref{F(t)})
is a strong solution of the Cauchy problem (\ref{NelBog1}),(\ref{eq:NelBog2}) and for
arbitrary initial data from the space $L^{1}_{\alpha}$ it is a weak solution \cite{GRS04}.

We remark, as a result of the application to cumulants (\ref{nLkymyl}) of analogs of
the Duhamel equations, solution series (\ref{F(t)}) reduces to the iteration series
of the BBGKY hierarchy with hard sphere collisions (\ref{NelBog1}).

In order to describe the evolution of infinitely many hard spheres we must construct
the solutions for initial data from more general Banach spaces. In the capacity
of such Banach space in \cite{CGP97}-\cite{Sp91},\cite{L} it was used the space
$L^{\infty}_\xi$ of sequences $f=(f_0,f_1,\ldots,f_n,\ldots)$ of continuous functions
$f_n(x_1,\ldots,x_n)$ defined on the phase space of hard spheres, that are symmetric
with respect to permutations of the arguments $x_1,\ldots,x_n$, equal to zero on the
set of the forbidden configurations $\mathbb{W}_n$ and equipped with the norm:
\begin{eqnarray*}
   &&\|f\|_{L^{\infty}_\xi}=\sup\limits_{n\geq0}\xi^{-n}
      \sup\limits_{x_1,\ldots,x_{n}}|f_n(x_1,\ldots,x_{n})|\exp(\frac{\beta}{2}\sum_{i=1}^n p_i^2),
\end{eqnarray*}
where $\beta>0$ and $\xi>0$ are some parameters.

In the space $L^{\infty}_\xi$ for series expansion (\ref{F(t)}) the following statement is true.
If $F(0)\in L^{\infty}_\xi$, every term of series expansion (\ref{F(t)}) exists and this series
converges uniformly on each compact almost everywhere for finite time interval. The sequence of
marginal distribution functions (\ref{F(t)}) is a unique weak solution of the Cauchy problem
(\ref{NelBog1}),(\ref{eq:NelBog2}) of the BBGKY hierarchy for hard spheres.

\subsection{On the Boltzmann--Grad asymptotic behavior}
To consider the conventional approach to the derivation of the Boltzmann kinetic equation with
hard sphere collisions from underlying dynamics \cite{PG90} (see also \cite{CGP97} and references
cited therein) we shall represent a solution of the Cauchy problem of the BBGKY hierarchy for hard
spheres in the form of the perturbation (iteration) series:
\begin{eqnarray}\label{BBGKYpert}
   &&\hskip-9mm F_{s}(t,x_1,\ldots,x_{s})=\sum\limits_{n=0}^{\infty}\epsilon^{2n}\int_0^tdt_{1}\ldots
       \int_0^{t_{n-1}}dt_{n}\int_{(\mathbb{R}^{3}\times\mathbb{R}^{3})^{n}}dx_{s+1}\ldots dx_{s+n}\,
       S_{s}(-t+t_{1})\\
   &&\times\sum\limits_{i_{1}=1}^{s}\mathcal{L}_{\mathrm{int}}^{\ast}(i_{1},s+1)
       S_{s+1}(-t_{1}+t_{2})\ldots\nonumber\\
   &&\times S_{s+n-1}(-t_{n}+t_{n})\sum\limits_{i_{n}=1}^{s+n-1}
       \mathcal{L}_{\mathrm{int}}^{\ast}(i_{n},s+n)
       S_{s+n}(-t_{n})F_{s+n}^{0,\epsilon},\quad s\geq1,\nonumber
\end{eqnarray}
where the notations from formula (\ref{aLint}) are used. If $F(0)\in L^{\infty}_\xi$, every
term of series expansion (\ref{BBGKYpert}) exists \cite{GP85},\cite{PG90} and the iteration
series converges uniformly on each compact almost everywhere for finite time interval:
$|t|<t_0(\beta,\xi)$. In paper \cite{GP85} for initial data close to equilibrium states,
namely locally perturbed equilibrium distribution functions, and in paper \cite{Ger} for
arbitrary initial data from the space $L^{\infty}_\xi$ in a one-dimensional hard sphere
(rod) system it were proved the existence of series expansion (\ref{BBGKYpert}) for arbitrary
time interval.

The sequence of marginal distribution functions (\ref{BBGKYpert}) is a unique weak solution
of the Cauchy problem (\ref{NelBog1}),(\ref{eq:NelBog2}) of the BBGKY hierarchy for hard
spheres \cite{GP85}.

Let the set $\mathbb{K}_s^0\subset\mathbb{R}^{3s}\setminus \mathbb{W}_s$ be an arbitrary
compact set of the admissible configurations of a system of $s$ hard spheres consisting from
the configurations: $|q_i-q_j|\geq\epsilon_0(\epsilon)+\epsilon, i\neq j\in(1,\ldots,s)$, where
$\epsilon_0(\epsilon)$ is a fixed value such that: $\lim_{\epsilon\rightarrow 0}\epsilon_0(\epsilon)=0$
and $\lim_{\epsilon\rightarrow 0}\frac{\epsilon}{\epsilon_0(\epsilon)}=0$.
Then the Boltzmann--Grad asymptotic behavior of perturbative solution (\ref{BBGKYpert}) is described
by the following statement \cite{PG89},\cite{PG90}.
\begin{theorem}
If for initial data $F(0)=(1,F_1^{0,\epsilon},\ldots,F_n^{0,\epsilon},\ldots)\in L^{\infty}_\xi$
uniformly on every compact set from the phase space
$\mathbb{R}^{3n}\times(\mathbb{R}^{3n}\setminus\mathbb{W}_n)$ it holds:
$\lim_{\epsilon\rightarrow 0}|\epsilon^{2n}F_{n}^{0,\epsilon}(x_1,\ldots,x_{n})-f_{n}^0(x_1,\ldots,x_{n})|=0,$
then for any finite time interval the function $\epsilon^{2s}F_{s}(t,x_1,\ldots,x_{s})$
defined by series (\ref{BBGKYpert}) converges in the Boltzmann--Grad limit uniformly with
respect to configuration variables from the set $\mathbb{K}_s^0$ and in a weak
sense with respect to momentum variables to the limit marginal distribution function
$f_s(t,x_1,\ldots,x_{s})$ given by the series expansion
\begin{eqnarray}\label{Iter2}
   &&\hskip-9mm f_{s}(t,x_1,\ldots,x_{s})=
        \sum\limits_{n=0}^{\infty}\int_0^tdt_{1}\ldots\hskip-2mm\int_0^{t_{n-1}}dt_{n}
        \int_{(\mathbb{R}^{3}\times\mathbb{R}^{3})^{n}}dx_{s+1}\ldots dx_{s+n}
        \prod\limits_{i_1=1}^{s}S_{1}(-t+t_{1},i_1)\\
   &&\times\sum\limits_{k_{1}=1}^{s}\mathcal{L}_{\mathrm{int}}^{0,\ast}(k_{1},s+1)
        \prod\limits_{j_1=1}^{s+1}S_{1}(-t_{1}+t_{2},j_1)\ldots \nonumber\\
   &&\times\prod\limits_{i_n=1}^{s+n-1} S_{1}(-t_{n}+t_{n},i_n)
         \sum\limits_{k_{n}=1}^{s+n-1}\mathcal{L}_{\mathrm{int}}^{0,\ast}(k_{n},s+n)
         \prod\limits_{j_n=1}^{s+n}S_{1}(-t_{n},j_n)f_{s+n}^0.\nonumber
\end{eqnarray}
\end{theorem}
In series expansion (\ref{Iter2}) the following operator is introduced:
\begin{eqnarray}\label{aLint0}
   &&\hskip-9mm(\mathcal{L}_{\mathrm{int}}^{0,\ast}(j_1,j_{2})f_n)(x_1,\ldots,x_n)\doteq
       \int_{\mathbb{S}_{+}^2}d\eta\langle\eta,(p_{j_1}-p_{j_2})\rangle \big(f_n(x_1,\ldots\\
   &&q_{j_1},p_{j_1}^\ast,\ldots q_{j_2},p_{j_2}^\ast,\ldots,x_n)-f_n(x_1,\ldots,x_n)\big)
       \delta(q_{j_1}-q_{j_2}),\nonumber
\end{eqnarray}
where the notations accepted in formula (\ref{aLint}) are used and the momenta
$p_{j_1}^{\ast},p_{j_2}^{\ast}$ are given by expressions (\ref{momenta}) as above.

If $f(0)=(1,f_1^0,\ldots,f_n^0,\ldots)\in L^{\infty}_\xi$, every term of series expansion
(\ref{Iter2}) exists and this series converges uniformly on each compact almost everywhere
for finite time interval $t<t_0(\beta,\xi)$.

We note that for $t\geq0$ sequence (\ref{Iter2}) is a weak solution of the Cauchy problem
of the limit BBGKY hierarchy known as the Boltzmann hierarchy with hard sphere collisions:
\begin{eqnarray}
  \label{Bl1}
    &&\hskip-9mm\frac{\partial}{\partial t}f_s(t)=\sum\limits_{j=1}^{n}\mathcal{L}^\ast(j)f_{s}(t)+
       \sum_{i=1}^{s}\,\int_{\mathbb{R}^{3}\times\mathbb{R}^{3}}dx_{s+1}
       \mathcal{L}^{0,\ast}_{\mathrm{int}}(i,s+1)f_{s+1}(t),\\ \nonumber\\
  \label{Bi2}
    &&\hskip-9mm f_s(t)_{\mid t=0}=f_s^0, \quad s\geq1,
\end{eqnarray}
where the operators $\mathcal{L}^\ast(j)$ and $\mathcal{L}^{0,\ast}_{\mathrm{int}}(i,s+1)$ are
defined by formulae (\ref{aL}) and (\ref{aLint0}), respectively.

We remark that the same statement takes place concerning the Boltzmann--Grad behavior of
non-perturbative solution (\ref{F(t)}) of the Cauchy problem of the BBGKY hierarchy for
hard spheres.

To derive the Boltzmann kinetic equation \cite{V} we will consider initial data (\ref{eq:NelBog2})
specified by a one-particle distribution function, namely initial data satisfying a chaos condition
\cite{CGP97}
\begin{eqnarray}\label{eq:Bog2_haos}
  &&\hskip-7mm F^{0,\epsilon}_{s}(x_1,\ldots,x_{s})
     =\prod_{i=1}^{s}F_{1}^{0,\epsilon}(x_i)
     \mathcal{X}_{\mathbb{R}^{3s}\setminus\mathbb{W}_s^{\epsilon}},\quad s\geq1,
\end{eqnarray}
where $\mathcal{X}_{\mathbb{R}^{3s}\setminus\mathbb{W}_s^{\epsilon}}$ is a characteristic
function of the set $\mathbb{R}^{3s}\setminus\mathbb{W}_s^{\epsilon}$ of allowed configurations.
Such assumption about initial data is intrinsic for the kinetic description of a gas,
because in this case all possible states are characterized only by a one-particle
marginal distribution function.

Since the initial limit marginal distribution functions satisfy a chaos condition too, i.e.
\begin{eqnarray}\label{lih2}
    &&\hskip-5mmf_s^0(x_1,\ldots,x_{s})=\prod \limits_{i=1}^{s}f_{1}^0(x_i), \quad s\geq2,
\end{eqnarray}
perturbative solution (\ref{Iter2}) of the Cauchy problem (\ref{Bl1}),(\ref{Bi2}) of the
Botzmann hierarchy has the following property (the propagation of initial chaos):
\begin{eqnarray*}
   &&\hskip-5mm f_{s}(t,x_1,\ldots,x_{s})=\prod\limits_{i=1}^{s}f_{1}(t,x_i), \quad s\geq2,
\end{eqnarray*}
where for $t\geq0$ the limit one-particle distribution function is determined by series
expansion (\ref{Iter2}) in case of $s=1$ and initial data (\ref{lih2}). If $t\geq0$, the
limit one-particle distribution function is governed by the Boltzmann kinetic equation with
hard sphere collisions \cite{CGP97}
\begin{eqnarray*}
    &&\hskip-8mm\frac{\partial}{\partial t}f_{1}(t,x_1)=
       -\langle p_1,\frac{\partial}{\partial q_1}\rangle f_{1}(t,x_1)+
       \int_{\mathbb{R}^3\times\mathbb{S}^2_+}d p_2\, d\eta\,\langle\eta,(p_1-p_2)\rangle\\
    &&\hskip+5mm\times\big(f_1(t,q_1,p_1^{\ast})f_1(t,q_1,p_2^{\ast})-
       f_1(t,x_1)f_1(t,q_1,p_2)\big), \nonumber
\end{eqnarray*}
where the momenta $p_{1}^{\ast}$ and $p_{2}^{\ast}$ are given by expressions (\ref{momenta}).

The Boltzmann--Grad asymptotic behavior of the equilibrium state of infinitely many hard spheres
and the systems of particles interacting via a short-range potential is established in paper \cite{PG88}
(see also \cite{CGP97}). It was proved that the equilibrium marginal distribution functions in the
Boltzmann--Grad scaling limit converge uniformly on any compacts to the Maxwell distribution functions.

\subsection{The dual BBGKY hierarchy for hard spheres}
Let $C_{\gamma}$ be the space of sequences $b=(b_0,b_1,\ldots,b_n,\ldots)$ of continuous functions
$b_n\in C_n$ equipped with the norm: $\|b\|_{C_{\gamma}}=\max_{n\geq 0}\,\frac{\gamma^{n}}{n!}\,
\|b_n\|_{C_n}=\max_{n\geq 0}\,\frac{\gamma^{n}}{n!}\sup_{x_1,\ldots,x_{n}}|b_n(x_1,\ldots,x_n)|$,
and $C_{\gamma}^0\subset C_{\gamma}$ is the set of finite sequences of infinitely differentiable
functions with compact supports.

If $t\geq0$, the evolution of marginal observables of a system of a non-fixed number of hard spheres
is described by the Cauchy problem of the weak formulation of the dual BBGKY hierarchy \cite{BGer}:
\begin{eqnarray}\label{dh}
   &&\hskip-9mm\frac{\partial}{\partial t}B_{s}(t)=\big(\sum\limits_{j=1}^{s}\mathcal{L}(j)+
      \epsilon^{2}\sum\limits_{j_1<j_{2}=1}^{s}\mathcal{L}_{\mathrm{int}}(j_1,j_{2})\big)B_{s}(t)+\\
   &&+\epsilon^{2}\sum_{j_1\neq j_{2}=1}^s
      \mathcal{L}_{\mathrm{int}}(j_1,j_{2})B_{s-1}(t,x_1,\ldots,x_{j_1-1},x_{j_1+1},\ldots,x_s),
      \nonumber\\
      \nonumber\\
  \label{dhi}
   &&\hskip-9mmB_{s}(t,x_1,\ldots,x_s)_{\mid t=0}=B_{s}^{\epsilon,0}(x_1,\ldots,x_s),\quad s\geq1.
\end{eqnarray}
In recurrence evolution equations (\ref{dh}) represented in a dimensionless form, as above,
the coefficient $\epsilon>0$ is a scaling parameter (the ratio of the diameter $\sigma>0$
to the mean free path of hard spheres) and operators (\ref{aL}) and (\ref{aLint}) are the adjoint
operators to the operators $\mathcal{L}(j)$ and $\mathcal{L}_{\mathrm{int}}(j_1,j_{2})$ defined
on $C_{s}^0$ as follows:
\begin{eqnarray}\label{com}
   &&\hskip-5mm\mathcal{L}(j)b_n\doteq\langle p_{j},\frac{\partial}{\partial q_{j}}\rangle b_n,\\
\label{comint}
   &&\hskip-5mm(\mathcal{L}_{\mathrm{int}}(j_1,j_{2})b_n)(x_1,\ldots,x_n)\doteq
     \int_{\mathbb{S}_+^2}d\eta\langle\eta,(p_{j_1}-p_{j_2})\rangle
     \big(b_n(x_1,\ldots,
\end{eqnarray}
\begin{eqnarray*}
   &&q_{j_1},p_{j_1}^\ast,\ldots,q_{j_2},p_{j_2}^\ast,\ldots,x_n)
     -b_n(x_1,\ldots,x_n)\big)\delta(q_{j_1}-q_{j_2}+\epsilon\eta),\nonumber
\end{eqnarray*}
where the  the post-collision momenta $p_{j_1}^\ast,p_{j_2}^\ast$ are determined by
expressions (\ref{momenta}) and $\mathbb{S}_{+}^{2}\doteq\{\eta\in\mathbb{R}^{3}
\big|\left|\eta\right|=1,\,\langle\eta,(p_{j_2}-p_{j_2})\rangle\geq0\}$.  If $t\leq0$, a generator of the dual BBGKY hierarchy
is determined by corresponding operator \cite{BGer}.

Let $Y\equiv(1,\ldots,s), Z\equiv (j_1,\ldots,j_{n})\subset Y$ and $\{Y\setminus Z\}$ is the
set consisting from one element $Y\setminus Z=(1,\ldots,j_{1}-1,j_{1}+1,\ldots,j_{n}-1,j_{n}+1,\ldots,s)$.

The non-perturbative solution $B(t)=(B_{0},B_{1}(t,x_1),\ldots,B_{s}(t,x_1,\ldots,x_s),\ldots)$
of the Cauchy problem (\ref{dh}),(\ref{dhi}) is represented by the sequence of marginal ($s$-particle)
observables:
\begin{eqnarray}\label{sdh}
   &&\hskip-12mm B_{s}(t,x_1,\ldots,x_s)=\sum_{n=0}^s\,
      \frac{1}{n!}\sum_{j_1\neq\ldots\neq j_{n}=1}^s
      \mathfrak{A}_{1+n}\big(t,\{Y\setminus Z\},Z\big) B_{s-n}^{\epsilon,0}(x_1,\\
   &&\ldots x_{j_1-1},x_{j_1+1},\ldots,x_{j_n-1},x_{j_n+1},\ldots,x_s), \quad s\geq1,\nonumber
\end{eqnarray}
where the generating operator of $n$ term of this expansion is the $(1+n)th$-order cumulant
of the groups of operators of hard spheres defined by the formula:
\begin{eqnarray}\label{cumulant}
    &&\hskip-5mm\mathfrak{A}_{1+n}(t,\{Y\setminus Z\},Z)\doteq
       \sum\limits_{\mathrm{P}:\,(\{Y\setminus Z\},Z)={\bigcup}_i X_i}
       (-1)^{\mathrm{|P|}-1}({\mathrm{|P|}-1})!\prod_{X_i\subset \mathrm{P}}
       S_{|\theta(X_i)|}(t,\theta(X_i)),
\end{eqnarray}
and notations accepted in formula (\ref{nLkymyl}) are used. The simplest examples of marginal
observables (\ref{sdh}) are given by the following expansions:
\begin{eqnarray*}
   &&B_{1}(t,x_1)=\mathfrak{A}_{1}(t,1)B_{1}^{\epsilon,0}(x_1),\\
   &&B_{2}(t,x_1,x_2)=\mathfrak{A}_{1}(t,\{1,2\})B_{2}^{\epsilon,0}(x_1,x_2)+
      \mathfrak{A}_{2}(t,1,2)(B_{1}^{\epsilon,0}(x_1)+B_{1}^{\epsilon,0}(x_2)).
\end{eqnarray*}

If $\gamma<e^{-1}$, then for $t\in\mathbb{R}$ in case of
$B(0)=(B_{0},B_{1}^{\epsilon,0},\ldots,B_{s}^{\epsilon,0},\ldots)\in C_{\gamma}^0\subset C_{\gamma}$
the sequence of marginal observables (\ref{sdh}) is a classical solution and
for arbitrary initial data $B(0)\in C_{\gamma}$ it is a generalized solution.

We remark that expansion (\ref{sdh}) can be also represented in the form of the perturbation
(iteration) series of the dual BBGKY hierarchy (\ref{dh}) as a result of applying of analogs
of the Duhamel equations to cumulants (\ref{cumulant}) of the groups of operators of hard spheres \cite{BGer}.

The single-component sequences of marginal observables correspond to observables of certain structure,
namely the marginal observables $b^{(1)}=(0,b_{1}^{(1)}(x_1),0,\ldots)$ correspond to the additive-type
observable, and the marginal observables $b^{(k)}=(0,\ldots,0,b_{k}^{(k)}(x_1,\ldots,x_k),0,\ldots)$
corresponds to the $k$-ary-type observables \cite{BGer}. If in capacity of initial data (\ref{dhi})
we consider the additive-type marginal observable, then the structure of solution expansion \eqref{sdh}
is simplified and attains the form
\begin{eqnarray}\label{af}
     &&\hskip-8mm B_{s}^{(1)}(t,x_1,\ldots,x_s)=
          \mathfrak{A}_{s}(t,1,\ldots,s)\sum_{j=1}^sB_{1}^{(1),\epsilon}(0,x_j), \quad s\geq 1.
\end{eqnarray}

We note that the mean value of the marginal observable $B(t)\in C_{\gamma}$ at
$t\in \mathbb{R}$ in the initial marginal state
$F(0)=(1,F_{1}^{\epsilon,0},\ldots,F_{n}^{\epsilon,0},\ldots)\in L^{1}=\bigoplus_{n=0}^{\infty}L^{1}_{n}$
is defined by the following functional:
\begin{eqnarray}\label{avmar-1}
   &&\hskip-5mm\big\langle B(t)\big|F(0)\big\rangle=\sum\limits_{s=0}^{\infty}\,
      \frac{1}{s!}\int_{(\mathbb{R}^{3}\times\mathbb{R}^{3})^{s}}
      dx_{1}\ldots dx_{s}\,B_{s}(t,x_1,\ldots,x_s)F_{s}^{\epsilon,0}(x_1,\ldots,x_s).
\end{eqnarray}
Owing to the estimate: $\|B(t)\|_{C_{\gamma}}\leq e^2(1-\gamma e)^{-1}\|B(0)\|_{C_{\gamma}}$,
functional (\ref{avmar-1}) exists under the condition that: $\gamma<e^{-1}$. In case of
$F(0)\in L^{\infty}_\xi$ the existence of mean value functional (\ref{avmar-1}) is proved
in a one-dimensional space in paper \cite{R10}.

In a special case functional (\ref{avmar-1}) of mean values of the additive-type marginal
observables $B^{(1)}(0)=(0,B_{1}^{(1),\epsilon}(0,x_1),0,\ldots)$  takes the form:
\begin{eqnarray*}\label{avmar-11}
   &&\hskip-5mm\big\langle B^{(1)}(t)\big|F(0)\big\rangle=
      \big\langle B^{(1)}(0)\big|F(t)\big\rangle=\int_{\mathbb{R}^{3}\times\mathbb{R}^{3}}dx_{1}\,
      B_{1}^{(1),\epsilon}(0,x_1)F_{1}(t,x_1),\nonumber
\end{eqnarray*}
where the sequence $B^{(1)}(t)$ is represented by (\ref{af}) and the one-particle distribution function
$F_{1}(t)$ is given by series expansion (\ref{F(t)}).

In the general case for mean values of marginal observables the following equality is true:
\begin{eqnarray}\label{eqos}
   &&\hskip-5mm\big\langle B(t)\big|F(0)\big\rangle=\big\langle B(0)\big|F(t)\big\rangle,
\end{eqnarray}
where the sequence $F(t)$ is given by formula (\ref{F(t)}). This equality signify the
equivalence of two pictures of the description of the evolution of hard spheres by means
of the BBGKY hierarchy (\ref{NelBog1}) and the dual BBGKY hierarchy for hard spheres (\ref{dh}).

\subsection{The generalized Enskog kinetic equation}
In paper \cite{GG} it was proved that, if initial states of a hard sphere system is
specified in terms of a one-particle distribution function on allowed configurations, then
at arbitrary moment of time the evolution of states governed by the BBGKY hierarchy can
be completely described within the framework of the one-particle marginal distribution
function $F_{1}(t)$ governed by the generalized Enskog kinetic equation. In this case all
possible correlations, generating by hard sphere dynamics, are described in terms of the
explicitly defined marginal functionals of the state $F_{s}\big(t\mid F_{1}(t)\big),\,s\geq2$.

If $t\geq 0$, the one-particle distribution function is governed by the Cauchy
problem of the following generalized Enskog equation \cite{GG}:
\begin{eqnarray}\label{gke1}
   &&\hskip-9mm\frac{\partial}{\partial t}F_{1}(t,x_1)=
       -\langle p_1,\frac{\partial}{\partial q_1}\rangle F_{1}(t,x_1)+
       \epsilon^2\int_{\mathbb {R}^3\!\times\mathbb{S}_{+}^{2}}
       d p_{2}d\eta\,\langle\eta,(p_1-p_{2})\rangle\times\\
   &&\times\big(F_{2}(t,q_1,p_1^\ast,q_1-\epsilon\eta,p_{2}^\ast\mid F_{1}(t))
       -F_{2}(t,x_1,q_1+\epsilon\eta,p_{2}\mid F_{1}(t))\big),\nonumber\\ \nonumber\\
 \label{2}
   &&\hskip-9mm F_1(t,x_1)_{\mid t=0}=F_1^{\epsilon,0}(x_1).
\end{eqnarray}
In kinetic equation (\ref{gke1}) represented in a dimensionless form, as above, the coefficient
$\epsilon>0$ is a scaling parameter, the pre-collision momenta $p_{1}^\ast,p_{2}^\ast$
are determined by expressions (\ref{momenta}), $\mathbb{S}_{+}^{2}\doteq\{\eta\in\mathbb{R}^{3}\big|\left|\eta\right|=1,\,\langle\eta,(p_{1}-p_{2})\rangle\geq0\}$,
and the collision integral is defined in terms of the marginal functional of the state in case of $s=2$.

The the marginal functionals of the state $F_{s}(t,x_1,\ldots,x_s\mid F_{1}(t)),\,s\geq 2,$
are represented by the following series expansion:
\begin{eqnarray}\label{f}
   &&\hskip-9mm F_{s}(t,x_1,\ldots,x_s\mid F_{1}(t))\doteq\\
   && \sum_{n=0}^{\infty}\frac{1}{n!}\,\int_{(\mathbb{R}^{3}\times\mathbb{R}^{3})^{n}}
     dx_{s+1}\ldots dx_{s+n}\,\mathfrak{V}_{1+n}(t,\{Y\},X\setminus Y)\prod_{i=1}^{s+n}F_{1}(t,x_i),\nonumber
\end{eqnarray}
where the notations accepted above are used: $Y\equiv(1,\ldots,s),\,X\equiv(1,\ldots,s+n)$, and
the $(n+1)th$-order generating evolution operator $\mathfrak{V}_{1+n}(t),\,n\geq0$, is defined by
the expansion:
\begin{eqnarray}\label{skrrn}
    &&\hskip-12mm\mathfrak{V}_{1+n}(t,\{Y\},X\setminus Y)\doteq\\
    &&\hskip-9mm \sum_{k=0}^{n}(-1)^k\,
       \sum_{m_1=1}^{n}\ldots\sum_{m_k=1}^{n-m_1-\ldots-m_{k-1}}\frac{n!}{(n-m_1-\ldots-m_k)!}
       \widehat{\mathfrak{A}}_{1+n-m_1-\ldots-m_k}(t)\nonumber\\
    &&\hskip-9mm \prod_{j=1}^k\,\sum_{k_2^j=0}^{m_j}\ldots
       \sum_{k^j_{n-m_1-\ldots-m_j+s}=0}^{k^j_{n-m_1-\ldots-m_j+s-1}}\prod_{i_j=1}^{s+n-m_1-\ldots-m_j}
       \frac{1}{(k^j_{n-m_1-\ldots-m_j+s+1-i_j}-k^j_{n-m_1-\ldots-m_j+s+2-i_j})!}\nonumber\\
    &&\hskip-9mm\times\widehat{\mathfrak{A}}_{1+k^j_{n-m_1-\ldots-m_j+s+1-i_j}-k^j_{n-m_1-\ldots-m_j+s+2-i_j}}
       (t,i_{j},s+n-m_1-\ldots-m_j+1+\nonumber \\
    &&\hskip-9mm +k^j_{s+n-m_1-\ldots-m_j+2-i_j},\ldots,s+n-m_1-\ldots-m_j+k^j_{s+n-m_1-\ldots-m_j+1-i_j}).\nonumber
\end{eqnarray}
In expression (\ref{skrrn}) it means that: $k^j_1\equiv m_j$ and $k^j_{n-m_1-\ldots-m_j+s+1}\equiv 0$
and we denote by the evolution operator $\widehat{\mathfrak{A}}_{1+n-m_1-\ldots-m_k}(t)
\equiv \widehat{\mathfrak{A}}_{1+n-m_1-\ldots-m_k}(t,\{Y\},s+1,\ldots,s+n-m_1-\ldots-m_k)$
the $(n-m_1-\ldots-m_k)th$-order scattering cumulant, namely
\begin{eqnarray*}\label{scacu}
   &&\hskip-8mm\widehat{\mathfrak{A}}_{1+n}(t,\{Y\},X\setminus Y)\doteq
       \mathfrak{A}_{1+n}(-t,\{Y\},X\setminus Y)
       \mathcal{X}_{\mathbb{R}^{3(s+n)}\setminus \mathbb{W}^{\epsilon}_{s+n}}
       \prod_{i=1}^{s+n}\mathfrak{A}_{1}(t,i), \quad n\geq1,\nonumber
\end{eqnarray*}
where the operator $\mathfrak{A}_{1+n}(-t)$ is $(1+n)th$-order cumulant
(\ref{nLkymyl}) of the adjoint groups of operators of hard spheres. We give
several simplest examples of generating evolution operators (\ref{skrrn}):
\begin{eqnarray*}
   &&\hskip-9mm\mathfrak{V}_{1}(t,\{Y\})=\widehat{\mathfrak{A}}_{1}(t,\{Y\})\doteq
       S_s(-t,1,\ldots,s)\mathcal{X}_{\mathbb{R}^{3s}\setminus \mathbb{W}^{\epsilon}_{s}}\prod_{i=1}^{s}S_1(t,i),\\
   &&\hskip-9mm\mathfrak{V}_{2}(t,\{Y\},s+1)=\widehat{\mathfrak{A}}_{2}(t,\{Y\},s+1)-
       \widehat{\mathfrak{A}}_{1}(t,\{Y\})\sum_{i_1=1}^s
       \widehat{\mathfrak{A}}_{2}(t,i_1,s+1).\nonumber
\end{eqnarray*}

If $\|F_{1}(t)\|_{L^{1}(\mathbb{R}^{3}\times\mathbb{R}^{3})}<e^{-(3s+2)}$, series expansion
\ref{f}) converges in the norm of the space $L^{1}_{s}$ for arbitrary $t\in\mathbb{R}$  \cite{GG},
and thus, the series expansion of the collision integral of kinetic equation (\ref{gke1})
converges under the condition that: $\|F_1(t)\|_{L^1(\mathbb{R}\times\mathbb{R})}<e^{-8}$.

A solution of the Cauchy problem (\ref{gke1}),(\ref{2}) is represented by the series expansion \cite{GG}:
\begin{eqnarray}\label{F(t)1}
    &&\hskip-12mm F_{1}(t,x_1)=\sum\limits_{n=0}^{\infty}\frac{1}{n!}
      \int_{(\mathbb{R}^3\times\mathbb{R}^3)^n}dx_2\ldots dx_{n+1}\,
      \mathfrak{A}_{1+n}(-t)\prod_{i=1}^{n+1}F_{1}^{\epsilon,0}(x_i)
      \mathcal{X}_{\mathbb{R}^{3(n+1)}\setminus \mathbb{W}^{\epsilon}_{n+1}},
\end{eqnarray}
where the generating operator $\mathfrak{A}_{1+n}(-t)\equiv\mathfrak{A}_{1+n}(-t,1,\ldots,n+1)$
is the $(1+n)th$-order cumulant (\ref{nLkymyl}) of adjoint groups of operators of hard spheres.
If initial one-particle distribution function $F_{1}^{\epsilon,0}$ is a continuously
differentiable integrable function with compact support, then function (\ref{F(t)1}) is a strong
solution of the Cauchy problem (\ref{gke1}),(\ref{2}) and for the arbitrary integrable function
$F_{1}^{\epsilon,0}$ it is a weak solution.

If initial one-particle marginal distribution function satisfies the following condition:
\begin{eqnarray}\label{G_1}
    &&|F_{1}^{\epsilon,0}(x_1)|\leq ce^{\textstyle-\frac{\beta}{2}{p^{2}_1}},
\end{eqnarray}
where $\beta>0$ is a parameter, $c<\infty$ is some constant, then every term of series
expansion (\ref{F(t)1}) exists, series (\ref{F(t)1}) converges uniformly on each compact almost
everywhere with respect to $x_1$ for finite time interval and function (\ref{F(t)1}) is a unique
weak solution of the Cauchy problem (\ref{gke1}),(\ref{2}) of the generalized Enskog kinetic equation.

The proof of the last statement is based on analogs of the Duhamel equations for cumulants
 (\ref{nLkymyl}) of groups of adjoint operators of hard spheres and the estimates established
 for the iteration series of the BBGKY hierarchy with hard sphere collisions \cite{CGP97}.

We point out the relationship of the description of the evolution of many hard spheres in terms
of the marginal observables and by the one-particle marginal distribution function governed by
the generalized Enskog kinetic equation (\ref{gke1}).

For mean value functional (\ref{avmar-1}) the following equality holds:
\begin{eqnarray}\label{w}
    &&\big\langle B(t)\big|F^c(0)\big\rangle=\big\langle B(0)\big|F(t\mid F_{1}(t))\big\rangle,
\end{eqnarray}
where the sequence $B(t)$ is a sequence of marginal observables defined by expansions (\ref{sdh}),
the sequence $F^c(0)=\big(1,F_{1}^{0,\epsilon}(x_1),\ldots,
\prod_{i=1}^{n}F_{1}^{0,\epsilon}(x_i)\mathcal{X}_{\mathbb{R}^{3n}\setminus\mathbb{W}^{\epsilon}_n}\big)$
is a sequence of initial marginal distribution functions and
$F(t\mid F_{1}(t))=\big(1,F_1(t),F_2(t\mid F_{1}(t)),\ldots,F_s(t\mid F_{1}(t))\big)$ is the
sequence which consists from solution expansion (\ref{F(t)1}) of the generalized Enskog kinetic equation
and marginal functionals of the state (\ref{f}).

In particular case of the $s$-ary initial marginal observable
$B^{(s)}(0)=(0,\ldots,0,B_{s}^{(s),\epsilon}(0,x_1,\ldots,$ $x_s),0,\ldots),\,s\geq2$,
equality (\ref{w}) takes the form
\begin{eqnarray*}\label{avmar-12}
   &&\hskip-9mm\big\langle B^{(s)}(t)\big|F^c(0)\big\rangle=
      \big\langle B^{(s)}(0)\big|F(t\mid F_{1}(t))\big\rangle=\\
   &&=\frac{1}{s!}\int_{(\mathbb{R}^{3}\times\mathbb{R}^{3})^{s}}dx_{1}\ldots dx_{s}\,
      B_{s}^{(s),\epsilon}(0,x_1,\ldots,x_s)F_s(t,x_1,\ldots,x_s\mid F_{1}(t)),\nonumber
\end{eqnarray*}
where the marginal functional of the state $F_{s}(t\mid F_{1}(t))$ is determined by series expansion
(\ref{f}). We emphasize that in fact functionals (\ref{f}) characterize the correlations
generated by dynamics of a hard sphere system with elastic collisions.

Correspondingly, in case of the additive-type marginal observable
$B^{(1)}(0)=(0,B_{1}^{(1),\epsilon}(0,x_1),$ $0,\ldots)$ equality (\ref{w}) takes the form
\begin{eqnarray*}\label{avmar-11}
   &&\hskip-5mm\big\langle B^{(1)}(t)\big|F^c(0)\big\rangle=
      \big\langle B^{(1)}(0)\big|F(t\mid F_{1}(t))\big\rangle=\int_{\mathbb{R}^{3}\times\mathbb{R}^{3}}dx_{1}\,
      B_{1}^{(1),\epsilon}(0,x_1)F_{1}(t,x_1),
\end{eqnarray*}
where the one-particle marginal distribution function $F_{1}(t)$ is represented by series expansion
(\ref{F(t)1}). Therefore for the additive-type marginal observables the generalized Enskog
kinetic equation (\ref{gke1}) is dual to the dual BBGKY hierarchy for hard spheres (\ref{dh})
with respect to bilinear form (\ref{avmar-1}).

Thus, if the initial state is specified by the one-particle distribution function
on allowed configurations, then the evolution of hard spheres governed by the dual BBGKY
hierarchy (\ref{dh}) for marginal observables of hard spheres can be completely described
within the framework of the generalized Enskog kinetic equation (\ref{gke1}) by the sequence
of marginal functionals of the state (\ref{f}).


\section{The kinetic evolution within the framework of marginal observables}
In this section we consider the problem of the rigorous description of the kinetic
evolution within the framework of many-particle dynamics of observables by giving
an example of the Boltzmann--Grad asymptotic behavior of a solution of the dual
BBGKY hierarchy with hard sphere collisions. Furthermore, we establish the links of
the dual Boltzmann hierarchy for the Boltzmann--Grad limit of marginal observables
with the Boltzmann kinetic equation.

\subsection{The Boltzmann--Grad asymptotics of the dual BBGKY hierarchy}
The Boltzmann--Grad scaling limit of non-perturbative solution (\ref{sdh}) of the Cauchy
problem (\ref{dh}),(\ref{dhi}) of the dual BBGKY hierarchy is described by the following
statement.
\begin{theorem}\label{3.1}
Let for $B_{n}^{\epsilon,0}\in C_n,\,n\geq1,$ it holds:
$\mathrm{w^{\ast}-}\lim_{\epsilon\rightarrow 0}(\epsilon^{-2n}B_{n}^{\epsilon,0}-b_{n}^0)=0,$
then for arbitrary finite time interval the Boltzmann--Grad limit of solution (\ref{sdh}) of
the Cauchy problem (\ref{dh}),(\ref{dhi}) of the dual BBGKY hierarchy exists in the
sense of the $\ast$-weak convergence on the space $C_s$
\begin{eqnarray}\label{asymto}
  &&\mathrm{w^{\ast}-}\lim\limits_{\epsilon\rightarrow 0}\big(\epsilon^{-2s}B_{s}(t)
        -b_{s}(t)\big)=0,
\end{eqnarray}
and it is determined by the following expansion:
\begin{eqnarray}\label{Iterd}
   &&\hskip-12mm b_{s}(t)=\sum\limits_{n=0}^{s-1}\,\int_0^tdt_{1}\ldots\int_0^{t_{n-1}}dt_{n}
      S_{s}^{0}(t-t_{1})\sum\limits_{i_{1}\neq j_{1}=1}^{s}\mathcal{L}_{\mathrm{int}}^0(i_{1},j_{1})
      S_{s-1}^{0}(t_{1}-t_{2})\ldots\\
   &&\hskip-8mm S_{s-n+1}^{0}(t_{n-1}-t_{n})
      \hskip-3mm\sum\limits^{s}_{\mbox{\scriptsize $\begin{array}{c}i_{n}\neq j_{n}=1,\\
      i_{n},j_{n}\neq (j_{1},\ldots,j_{n-1})\end{array}$}}\hskip-4mm
      \mathcal{L}_{\mathrm{int}}^0(i_{n},j_{n})S_{s-n}^{0}(t_{n})b_{s-n}^0((x_1,\ldots,x_s)
      \setminus(x_{j_{1}},\ldots,x_{j_{n}})),\nonumber\\
   &&\hskip-8mm s\geq1.\nonumber
\end{eqnarray}
\end{theorem}
In expansion (\ref{Iterd}) for groups of operators of noninteracting particles
the following notations are used:
\begin{eqnarray*}
   &&S_{s-n+1}^{0}(t_{n-1}-t_{n})\equiv S_{s-n+1}^{0}(t_{n-1}-t_{n},Y \setminus (j_{1},\ldots,j_{n-1}))=
     \prod\limits_{j\in Y \setminus (j_{1},\ldots,j_{n-1})}S_{1}(t_{n-1}-t_{n},j),
\end{eqnarray*}
and we denote by $\mathcal{L}_{\mathrm{int}}^{0}(j_1,j_{2})$ the operator:
\begin{eqnarray}\label{int0}
   &&\hskip-9mm(\mathcal{L}_{\mathrm{int}}^{0}(j_1,j_{2})b_n)(x_1,\ldots,x_n)\doteq
     \int_{\mathbb{S}_+^2}d\eta\langle\eta,(p_{j_1}-p_{j_2})\rangle
     \big(b_n(x_1,\ldots,q_{j_1},p_{j_1}^\ast,\ldots,\\
   &&\hskip+5mm q_{j_2},p_{j_2}^\ast,\ldots,x_n)-b_n(x_1,\ldots,x_n)\big)\delta(q_{j_1}-q_{j_2}),\nonumber
\end{eqnarray}
where
${\Bbb S}_{+}^{2}\doteq\{\eta\in\mathbb{R}^{3}\big|\,|\eta|=1,\langle\eta,(p_{j_1}-p_{j_2})\rangle\geq0\}$
and the momenta $p_{j_1}^{\ast},p_{j_2}^{\ast}$ are determined by expressions (\ref{momenta}).

Before to consider the proof scheme of the theorem we give some comments.
If $b^0\in C_{\gamma}$, then the sequence $b(t)=(b_0,b_1(t),\ldots,b_{s}(t),\ldots)$ of limit
marginal observables (\ref{Iterd}) is a generalized global solution of the Cauchy problem of
the dual Boltzmann hierarchy with hard sphere collisions
\begin{eqnarray}\label{vdh}
  &&\hskip-9mm \frac{d}{dt}b_{s}(t,x_1,\ldots,x_s)=
     \sum\limits_{j=1}^{s}\mathcal{L}(j)\,b_{s}(t,x_1,\ldots,x_s)+\\
  &&+\sum_{j_1\neq j_{2}=1}^s\mathcal{L}_{\mathrm{int}}^{0}(j_1,j_{2})
     b_{s-1}(t,x_1,\ldots,x_{j_1-1},x_{j_1+1},\ldots,x_s),\nonumber\\ \nonumber\\
 \label{vdhi}
   &&\hskip-9mm b_{s}(t,x_1,\ldots,x_s)_{\mid t=0}=b_{s}^0(x_1,\ldots,x_s), \quad s\geq1.
\end{eqnarray}
It should be noted that equations set (\ref{vdh}) has the structure of recurrence evolution equations.
We give several examples of the evolution equations of the dual Boltzmann hierarchy (\ref{vdh})
\begin{eqnarray*}
    &&\hskip-9mm\frac{\partial}{\partial t}b_{1}(t,x_1)=
       \langle p_1,\frac{\partial}{\partial q_{1}}\rangle\,b_{1}(t,x_1),\\
    &&\hskip-9mm\frac{\partial}{\partial t}b_{2}(t,x_1,x_2)=
       \sum\limits_{j=1}^{2}\langle p_j,\frac{\partial}{\partial q_{j}}\rangle\,b_{2}(t,x_1,x_2)+\int_{\mathbb{S}_{+}^2}d\eta\langle\eta,(p_{1}-p_{2})\rangle\\
    &&\times\big(b_1(t,q_{1},p_{1}^\ast)-b_1(t,x_1)+b_1(t,q_{2},p_{2}^\ast)-b_1(t,x_2)\big)\delta(q_{1}-q_{2}).
\end{eqnarray*}

The proof of the limit theorem for the dual BBGKY hierarchy is based on formulas for cumulants
of asymptotically perturbed groups of operators of hard spheres.

For arbitrary finite time interval the asymptotically perturbed group of operators of hard 
spheres has the following scaling limit in the sense of the $\ast$-weak convergence on the 
space $C_s$:
\begin{eqnarray}\label{Kato}
    &&\mathrm{w^{\ast}-}\lim\limits_{\epsilon\rightarrow 0}\big(S_s(t)b_s-
       \prod\limits_{j=1}^{s}S_{1}(t,j)b_s\big)=0.
\end{eqnarray}
Taking into account analogs of the Duhamel equations for cumulants of asymptotically
perturbed groups of operators, in view of formula (\ref{Kato}) we have
\begin{eqnarray*}\label{apc}
   &&\hskip-8mm \mathrm{w^{\ast}-}\lim\limits_{\epsilon\rightarrow 0}
       \Big(\epsilon^{-2n}\frac{1}{n!}
       \mathfrak{A}_{1+n}\big(t,\{Y\setminus X\},j_1,\ldots,j_{n}\big)\,b_{s-n}-\\
   && -\int_0^tdt_{1}\ldots\int_0^{t_{n-1}}dt_{n}
       \, S_{s}^{0}(t-t_{1})\sum\limits^{s}_{\mbox{\scriptsize $\begin{array}{c}i_{1}=1,\\
       i_{1}\neq j_{1}\end{array}$}}\mathcal{L}_{\mathrm{int}}^0(i_{1},j_{1})\,
       S_{s-1}(t_{1}-t_{2})\ldots\\
   && S_{s-n+1}^{0}(t_{n-1}-t_{n})
       \sum\limits^{s}_{\mbox{\scriptsize $\begin{array}{c}i_{n}=1,\\
       i_{n}\neq (j_{1},\ldots,j_{n})\end{array}$}}\mathcal{L}_{\mathrm{int}}^0(i_{n},j_{n})
       S_{s-n}^{0}(t_{n})\,b_{s-n}\Big)=0,
\end{eqnarray*}
where we used notations accepted in formula (\ref{Iterd}) and
$b_{s-n}\equiv b_{s-n}((x_1,\ldots,x_s)\setminus(x_{j_{1}},\ldots,x_{j_{n}}))$. As a result 
of this equality we establish the validity of statement (\ref{asymto}) for solution expansion
(\ref{sdh}) of the dual BBGKY hierarchy with hard sphere collisions (\ref{dh}).

We consider the Boltzmann--Grad limit of a special case of marginal observables, namely the 
additive-type marginal observables. As it was noted above in this case solution (\ref{sdh})
of the dual BBGKY hierarchy (\ref{dh}) is represented by formula (\ref{af}).

If for the initial additive-type marginal observable $B_{1}^{(1),\epsilon}(0)$ the following 
condition is satisfied:
\begin{eqnarray*}
    &&\hskip-5mm \mathrm{w^{\ast}-}\lim\limits_{\epsilon\rightarrow 0}\big( \epsilon^{-2}
        B_{1}^{(1),\epsilon}(0)-b_{1}^{(1)}(0)\big)=0,
\end{eqnarray*}
then, according to statement (\ref{asymto}), for additive-type marginal observable (\ref{af})
we derive
\begin{eqnarray*}
    &&\hskip-5mm \mathrm{w^{\ast}-}\lim\limits_{\epsilon\rightarrow 0}
                 \big(\epsilon^{-2s}B_{s}^{(1),\epsilon}(t)-b_{s}^{(1)}(t)\big)=0,
\end{eqnarray*}
where the limit marginal observable $b_{s}^{(1)}(t)$ is determined as a special case
of expansion (\ref{Iterd}):
\begin{eqnarray}\label{itvad}
   &&\hskip-10mm b_{s}^{(1)}(t,x_1,\ldots,x_s)=\int_0^t dt_{1}\ldots\int_0^{t_{s-2}}dt_{s-1}\,
       S_{s}^{0}(t-t_{1})\sum\limits_{i_{1}\neq j_{1}=1}^{s}
       \mathcal{L}_{\mathrm{int}}^0(i_{1},j_{1})\\
   &&\times S_{s-1}^{0}(t_{1}-t_{2})\ldots S_{2}^{0}(t_{s-2}-t_{s-1})\hskip-5mm
       \sum\limits^{s}_{\mbox{\scriptsize $\begin{array}{c}i_{s-1}\neq j_{s-1}=1,\\
       i_{s-1},j_{s-1}\neq (j_{1},\ldots,j_{s-2})\end{array}$}}\hskip-5mm
       \mathcal{L}_{\mathrm{int}}^0(i_{s-1},j_{s-1})\nonumber\\
   &&\times S_{1}^{0}(t_{s-1})\,b_{1}^{(1)}(0,(x_1,\ldots,x_s)
       \setminus (x_{j_{1}},\ldots,x_{j_{s-1}})),\quad s\geq1.\nonumber
\end{eqnarray}
We make several examples of expansions (\ref{itvad}) of the limit additive-type marginal 
observable:
\begin{eqnarray*}
   &&\hskip-8mm b_{1}^{(1)}(t,x_1)=S_{1}(t,1)\,b_{1}^{(1)}(0,x_1),\\
   &&\hskip-8mm b_{2}^{(1)}(t,x_1,x_2)=\int_0^t dt_{1}\prod\limits_{i=1}^{2}S_{1}(t-t_{1},i)\,
      \mathcal{L}_{\mathrm{int}}^0(1,2)\sum\limits_{j=1}^{2}S_{1}(t_{1},j)\,b_{1}^{(1)}(0,x_j).
\end{eqnarray*}

Thus, in the Boltzmann--Grad scaling limit the kinetic evolution of hard spheres
is described in terms of limit marginal observables (\ref{Iterd}) governed by the
dual Boltzmann hierarchy (\ref{vdh}). Similar approach to the description of the
mean field asymptotic behavior of quantum many-particle systems was developed in
paper \cite{G11}.

\subsection{The derivation of the Boltzmann kinetic equation}
We consider links of the constructed Boltzmann--Grad asymptotic behavior of the additive-type
marginal observables with the nonlinear Boltzmann kinetic equation. Furthermore, the relations
between the evolution of observables and the description of the kinetic evolution of states in
terms of a one-particle marginal distribution function are discussed.

Indeed, for the additive-type marginal observables the Boltzmann--Grad scaling limit gives an
equivalent approach to the description of the kinetic evolution of hard spheres in terms of the
Cauchy problem of the Boltzmann equation with respect to the Cauchy problem (\ref{vdh}),(\ref{vdhi})
of the dual Boltzmann hierarchy. In the case of the $k$-ary  marginal observable a solution of the
dual Boltzmann hierarchy (\ref{vdh}) is equivalent to the property of the propagation of initial
chaos for the $k$-particle marginal distribution function in the sense of equality (\ref{eqos}).

If $b(t)\in C_{\gamma}$ and $f_1^0\in L^{1}(\mathbb{R}^{3}\times\mathbb{R}^{3})$, then
under the condition that: $\|f_1^0\|_{L^{1}(\mathbb{R}^{3}\times\mathbb{R}^{3})}<\gamma$,
there exists the Boltzmann--Grad limit of mean value functional (\ref{avmar-1}) which is
determined by the series expansion
\begin{eqnarray*}
  &&\hskip-7mm \big\langle b(t)\big|f^{(c)}\big\rangle=\sum\limits_{s=0}^{\infty}\,\frac{1}{s!}\,
      \int_{(\mathbb{R}^{3}\times\mathbb{R}^{3})^{s}}
      dx_{1}\ldots dx_{s}\,b_{s}(t,x_1,\ldots,x_s)\prod\limits_{i=1}^{s} f_1^0(x_i),
\end{eqnarray*}
where we assumed that the initial state is specified by a one-particle marginal distribution 
function, namely it represents by the sequence
$f^{(c)}\equiv(1,f_1^0(\textbf{u}_1),\ldots,{\prod\limits}_{i=1}^{s}f_{1}^0(\textbf{u}_i),\ldots)$.

Consequently, for the limit additive-type marginal observables (\ref{itvad}) the following
equality is true:
\begin{eqnarray*}\label{avmar-2}
  &&\hskip-7mm \big\langle b^{(1)}(t)\big|f^{(c)}\big\rangle=
     \sum\limits_{s=0}^{\infty}\,\frac{1}{s!}\,\int_{(\mathbb{R}^{3}\times\mathbb{R}^{3})^{s}}
     dx_{1}\ldots dx_{s}\,b_{s}^{(1)}(t,x_1,\ldots,x_s)\prod \limits_{i=1}^{s} f_{1}^0(x_i)=\nonumber\\
  &&\hskip+7mm=\int_{\mathbb{R}^{3}\times\mathbb{R}^{3}}dx_{1}\,b_{1}^{(1)}(0,x_1)f_{1}(t,x_1),\nonumber
\end{eqnarray*}
where the function $b_{s}^{(1)}(t)$ is represented by expansion (\ref{itvad}) and the limit
marginal distribution function $f_{1}(t,x_1)$ is represented by the series expansion
\begin{eqnarray}\label{viter}
   &&\hskip-9mm f_{1}(t,x_1)=\sum\limits_{n=0}^{\infty}\int_0^tdt_{1}\ldots\int_0^{t_{n-1}}dt_{n}\,
        \int_{(\mathbb{R}^{3}\times\mathbb{R}^{3})^{n}}dx_{2}\ldots dx_{n+1}S_{1}(-t+t_{1},1)\\
   &&\times\mathcal{L}_{\mathrm{int}}^{0,\ast}(1,2)
        \prod\limits_{j_1=1}^{2}S_{1}(-t_{1}+t_{2},j_1)\ldots\nonumber\\
   &&\times\prod\limits_{i_{n}=1}^{n}S_{1}(-t_{n}+t_{n},i_{n})
        \sum\limits_{k_{n}=1}^{n}\mathcal{L}_{\mathrm{int}}^{0,\ast}(k_{n},n+1)
        \prod\limits_{j_n=1}^{n+1}S_{1}(-t_{n},j_n)\prod\limits_{i=1}^{n+1}f_1^0(x_i).\nonumber
\end{eqnarray}
In series (\ref{viter}) the operator (\ref{aLint0}) adjoint to operator (\ref{int0}) in the 
sense of functional (\ref{avmar-1}) is used.

If the function $f_1^0$ is continuous, every term of series expansion (\ref{viter}) exists and 
this series converges uniformly on each compact almost everywhere for finite time interval.

For $t\geq0$ limit marginal distribution function (\ref{viter}) is a weak solution of the Cauchy
problem of the Boltzmann kinetic equation with hard sphere collisions:
\begin{eqnarray}
  \label{Bolz}
    &&\hskip-5mm\frac{\partial}{\partial t}f_{1}(t,x_1)=
       -\langle p_1,\frac{\partial}{\partial q_1}\rangle f_{1}(t,x_1)+\\
    &&\hskip+5mm+\int_{\mathbb{R}^3\times\mathbb{S}^2_+}d p_2\,d\eta\,
      \langle\eta,(p_1-p_2)\rangle\big(f_1(t,q_1,p_1^{\ast})f_1(t,q_1,p_2^{\ast})-
       f_1(t,x_1)f_1(t,q_1,p_2)\big), \nonumber\\ \nonumber\\
  \label{Bolzi}
    &&\hskip-5mm f_1(t,x_1)_{\mid t=0}=f_{1}^0(x_1),
\end{eqnarray}
where the momenta $p_{1}^{\ast}$ and $p_{2}^{\ast}$ are determined by expressions (\ref{momenta}).

Thus, we establish that the dual Boltzmann hierarchy with hard sphere collisions (\ref{vdh})
for additive-type marginal observables and initial states specified by one-particle marginal
distribution function (\ref{lih2}) describes the evolution of a hard sphere system
just as the Boltzmann kinetic equation with hard sphere collisions (\ref{Bolz}).

\subsection{On the propagation of initial chaos}
We prove that within the framework of the evolution of marginal observables of hard
spheres in the Boltzmann--Grad scaling limit a chaos property for states is fulfilled
at arbitrary instant.

The property of the propagation of initial chaos is a consequence of the validity
of the following equality for the mean value functionals of the limit $k$-ary marginal
observables in the case of $k\geq2$:
\begin{eqnarray}\label{dchaos}
    &&\hskip-7mm \big\langle b^{(k)}(t)\big|f^{(c)}\big\rangle=\sum\limits_{s=0}^{\infty}\,\frac{1}{s!}\,
       \int_{(\mathbb{R}^{3}\times\mathbb{R}^{3})^{s}}dx_{1}\ldots dx_{s}\,b_{s}^{(k)}(t,x_1,\ldots,x_s)
       \prod \limits_{i=1}^{s} f_1^0(x_i)\\
    &&\hskip+7mm =\frac{1}{k!}\int_{(\mathbb{R}^{3}\times\mathbb{R}^{3})^{k}}
       dx_{1}\ldots dx_{k}\,b_{k}^{(k)}(0,x_1,\ldots,x_k)
       \prod\limits_{i=1}^{k} f_{1}(t,x_i),\nonumber
\end{eqnarray}
where the limit one-particle distribution function $f_{1}(t,x_i)$ is defined by series
expansion (\ref{viter}) and therefore it is governed by the Cauchy problem
(\ref{Bolz}),(\ref{Bolzi}) of the Boltzmann kinetic equation with hard sphere collisions.

Thus, in the Boltzmann--Grad scaling limit an equivalent approach to the description
of the kinetic evolution of hard spheres in terms of the Cauchy problem (\ref{Bolz}),(\ref{Bolzi})
of the Boltzmann kinetic equation is given by the Cauchy problem (\ref{vdh}),(\ref{vdhi})
of the dual Boltzmann hierarchy for the additive-type marginal observables.
In the general case of the $k$-ary marginal observables a solution of the dual Boltzmann
hierarchy (\ref{vdh}) is equivalent to the validity of a chaos property for the $k$-particle
marginal distribution functions in the sense of equality (\ref{dchaos}) or in other words
the Boltzmann--Grad scaling dynamics does not create correlations.

\section{The Boltzmann--Grad asymptotics of the generalized Enskog equation}
In this section we consider an approach to the rigorous derivation of the Boltzmann
equation with hard sphere collisions from the generalized Enskog kinetic equation.

\subsection{The Boltzmann--Grad limit theorem}
For a solution of the generalized Enskog kinetic equation (\ref{gke1}) the following
Boltzmann--Grad scaling limit theorem is true \cite{GG04}.
\begin{theorem}
If the initial one-particle marginal distribution function $F_{1}^{\epsilon,0}$ satisfies
condition (\ref{G_1}) and there exists its limit in the sense of a weak convergence:
$\mathrm{w-}\lim_{\epsilon\rightarrow 0}(\epsilon^{2}F_{1}^{\epsilon,0}(x_1)-f_{1}^0(x_1))=0,$
then for finite time interval the Boltzmann--Grad limit of solution (\ref{F(t)1}) of the Cauchy
problem (\ref{gke1}),(\ref{2}) of the generalized Enskog equation exists in the same sense
\begin{eqnarray}\label{asymt}
  &&\mathrm{w-}\lim\limits_{\epsilon\rightarrow 0}\big(\epsilon^{2}F_{1}(t,x_1)-f_{1}(t,x_1)\big)=0,
\end{eqnarray}
where the limit one-particle distribution function is represented by uniformly convergent
on arbitrary compact set series expansion (\ref{viter}).

If $f_{1}^0$ satisfies condition (\ref{G_1}), then for $t\geq 0$ the limit one-particle
distribution function represented by series (\ref{viter}) is a weak solution of the Cauchy
problem (\ref{Bolz}),(\ref{Bolzi}) of the Boltzmann kinetic equation with hard sphere collisions.
\end{theorem}

The proof of this theorem is based on formulas for cumulants (\ref{nLkymyl}) of asymptotically
perturbed groups of adjoint operators of hard spheres. Namely, in the sense of a weak convergence
the following equality holds:
\begin{eqnarray*}\label{lemma2}
    &&\hskip-5mm\mathrm{w-}\lim\limits_{\epsilon\rightarrow 0}\Big(\epsilon^{-2n}\frac{1}{n!}
        \mathfrak{A}_{1+n}(-t,1,\ldots,n+1)f_{n+1}-\\
    &&\hskip-5mm -\int_0^tdt_{1}\ldots\int_0^{t_{n-1}}dt_{n}S_{1}(-t+t_{1},1)
        \mathcal{L}_{\mathrm{int}}^{0,\ast}(1,2)\prod\limits_{j_1=1}^{2}S_{1}(-t_{1}+t_{2},j_1)\ldots\\
    &&\hskip-5mm\prod\limits_{i_{n}=1}^{n}S_{1}(-t_{n-1}+t_{n},i_{n})
        \sum\limits_{k_{n}=1}^{n}\mathcal{L}_{\mathrm{int}}^{0,\ast}(k_{n},n+1)
        \prod\limits_{j_n=1}^{n+1}S_{1}(-t_{n},j_n)f_{n+1}\Big)=0, \quad n\geq1,
\end{eqnarray*}
where notations accepted in formula (\ref{viter}) are used.

Thus, the Boltzmann--Grad scaling limit of solution (\ref{F(t)1}) of the generalized Enskog
equation is governed by the Boltzmann kinetic equation with hard sphere collisions (\ref{Bolz}).

We note that one of the advantage of the developed approach to the derivation of the Boltzmann
equation is the possibility to construct of the higher-order corrections to the Boltzmann--Grad
evolution of many-particle systems with hard sphere collisions.

\subsection{A scaling limit of marginal functionals of the state}
As we note above the all possible correlations of a hard sphere system are described
by marginal functionals of the state (\ref{f}). Taking into consideration the fact of
the existence of the Boltzmann--Grad scaling limit (\ref{asymt}) of solution  (\ref{F(t)1})
of the generalized Enskog kinetic equation (\ref{gke1}), for marginal functionals of the
state (\ref{f}) the following statement holds.
\begin{theorem}
Under the conditions of the Boltzmann--Grad limit theorem for the generalized Enskog kinetic
equation for finite time interval the following Boltzmann--Grad limits of marginal functionals
of the state (\ref{f}) exist in the sense of a weak convergence on the space of bounded functions:
\begin{eqnarray*}
   &&\mathrm{w-}\lim\limits_{\epsilon\rightarrow 0}\big(\epsilon^{2s}
      F_{s}\big(t,x_1,\ldots,x_s\mid F_{1}(t)\big)-\prod\limits_{j=1}^{s}f_{1}(t,x_j)\big)=0,
\end{eqnarray*}
where the limit one-particle distribution function $f_{1}(t)$ is represented
by series expansion (\ref{viter}).
\end{theorem}

The proof of this limit theorem is based on the validity of the following formulas for generating
evolution operators (\ref{skrrn}) of scattering cumulants of asymptotically perturbed adjoint groups
of operators of hard spheres:
\begin{eqnarray*}
    &&\mathrm{w-}\lim\limits_{\epsilon\rightarrow 0}\big(\mathfrak{V}_{1}(t,\{Y\})f_{s}-If_{s}\big)=0,\\
    &&\mathrm{w-}\lim\limits_{\epsilon\rightarrow 0}\epsilon^{-2n}
        \mathfrak{V}_{1+n}(t,\{Y\},X\setminus Y)f_{s+n}=0, \quad n\geq 1,
\end{eqnarray*}
where for $f=(1,f_1,\ldots,f_n\ldots)\in L^{\infty}_\xi$ these limits exist in the sense of
a weak convergence.

Thus, the Boltzmann--Grad scaling limits of marginal functionals of the state (\ref{f}) are
the products of solution (\ref{viter}) of the Boltzmann equation with hard sphere collisions
(\ref{Bolz}) that means the property of the propagation of initial chaos \cite{CGP97},\cite{CIP}.

\subsection{Remark: a one-dimensional system of hard spheres}
We consider the Boltzmann--Grad asymptotic behavior of a solution of the generalized
Enskog equation in a one-dimensional space. In this case the dimensionless collision
integral $\mathcal{I}_{GEE}$ has the structure \cite{Ger},\cite{GP83}:
\begin{eqnarray*}
  &&\hskip-9mm\mathcal{I}_{GEE}=\int_0^\infty dP\,P \big(F_{2}(t,q_1,p_1-P,
     q_1-\epsilon,p_1\mid F_{1}(t))-F_{2}(t,q_1,p_1,q_1-\epsilon,p_1+P\mid F_{1}(t)+\\
  &&+F_{2}(t,q_1,p_1+P,q_1+ \epsilon,p_1\mid F_{1}(t))
     -F_{2}(t,q_1,p_1,q_1+\epsilon,p_1-P\mid F_{1}(t))\big),\nonumber
\end{eqnarray*}
where $\epsilon>0$ is a scaling parameter and the marginal functional of the state
$F_{2}(t\mid F_{1}(t))$ is represented by series expansion (\ref{f}) in the case of $s=2$
in a one-dimensional space.

As we can see in the Boltzmann--Grad limit the collision integral of the generalized
Enskog equation in a one-dimensional space vanishes, i.e. in other words dynamics of
a one-dimensional system of elastically interacting hard spheres is trivial (a free
molecular motion or the Knudsen flow).

We note that in paper \cite{BG12} it was established that the Boltzmann--Grad asymptotic
behavior of inelastically interacting hard rods is not trivial in contrast to a one-dimensional
hard rod system with elastic collisions and it is governed by the Boltzmann-type kinetic equation
for granular gases.


\section{Conclusion and outlook}
In the paper two new approaches to the description of the kinetic evolution of
many-particle systems with hard sphere collisions were developed. One of them is
a formalism for the description of the evolution of infinitely many hard spheres
within the framework of marginal observables in the Boltzmann--Grad scaling limit.
Another approach to the description of the kinetic evolution of hard spheres is
based on the non-Markovian generalization of the Enskog kinetic equation.

In particular, it was established that a chaos property of the Boltzmann--Grad
scaling behavior of the $s$-particle marginal distribution function of infinitely
many hard spheres is equivalent in the sense of equality (\ref{dchaos}) to a solution
of the dual Boltzmann hierarchy (\ref{vdh}) in the case of the $s$-ary marginal observable.
In other words the Boltzmann--Grad scaling dynamics does not create correlations.

One of the advantage of the considered approaches is the possibility to construct
the kinetic equations in scaling limits, involving correlations at initial time,
which can characterize the condensed states of a hard sphere system.

We emphasize that the approach to the derivation of the Boltzmann equation from
underlying dynamics governed by the generalized Enskog kinetic equation enables
to construct also the higher-order corrections to the Boltzmann--Grad evolution
of many-particle systems with hard sphere collisions.

\addcontentsline{toc}{section}{References}
{\small
\renewcommand{\refname}{References}

\end{document}